\documentclass[cits]{PoS}
\pdfoutput=1
\usepackage{amsmath,amssymb,amsthm,latexsym}
\usepackage{stmaryrd,wasysym,upgreek,mathrsfs,dsfont}
\usepackage{graphicx,color}
\usepackage{physics}

\usepackage{tensor}
\usepackage{verbatim}
\usepackage{epigraph}
\usepackage{url}
\usepackage{footmisc}
\usepackage{physics}
\usepackage{todonotes}
\usepackage{wasysym} % smiley commands
\usepackage{marvosym} % Smiley commands
\usepackage{textcomp} % G

\newcommand{\beq}{\begin{equation}}
\newcommand{\eeq}{\end{equation}}
\newcommand{\be}{\begin{equation}}
\newcommand{\bee}{\begin{equation}}
\newcommand{\ee}{\end{equation}}
\newcommand{\bea}{\begin{eqnarray}}
\newcommand{\eea}{\end{eqnarray}}

\newtheorem{prop}{Proposition}[section]

\newtheorem{theorem}[prop]{Theorem}

\newcommand{\f}{\frac}

\newtheorem{lemma}{Lemma}[section]

\newcommand{\bal}{\begin{align}}
\newcommand{\eal}{\end{align}}

\newcommand{\cC}{{\cal{C}}}

\newcommand{\cT}{{\cal{T}}}

\newcommand{\cU}{{\cal{U}}}

\newcommand{\N}{\mathbb{N}}

\newcommand{\bbbone}{{\mathds{1}}}

\newcommand{\lam}{{\lambda}}
\newcommand{\mP}{{\mathbb{P}}}

\newcommand{\mE}{{\mathbb{E}}}

\title{The Tensor Track VI: \\
Field Theory on Random Trees and\\
SYK on Random Unicyclic Graphs}

\ShortTitle{Field Theory on Random Trees and Unicycles}

\author{Nicolas Delporte and \speaker{Vincent Rivasseau}\\
        Université Paris-Saclay, CNRS/IN2P3, IJCLab, 91405 Orsay, France\\
        CNRS UMR 9012 \\        E-mails: nicolas.delporte@ijclab.in2p3.fr, vincent.rivasseau@u-psud.fr}

\abstract{We review the paper \cite{DeRi} and discuss some comments about SYK on random unicycle.}

\FullConference{Corfu Summer Institute 2019 "School and Workshops on Elementary Particle Physics and Gravity" (CORFU2019)\\
		31 August - 25 September 2019\\
		Corfu, Greece}

\begin{document}
\tableofcontents

\section{Introduction}

The tensor track \cite{tensortrack} is an attempt to quantize gravity by making use of random tensor models \cite{tensors}. It lies at the crossroad of  several  closely  related  approaches,  most  notably  random  matrix  models,  (causal)  dynamical  triangulations,  non -commutative quantum field theory, and group field theory,  which is the second-quantized field-theoretic version of  loop  quantum  gravity.   

The  world  of  random  tensors models has  a  surprisingly  simple  entrance  door, namely the family of melonic graphs \cite{Bon}.  Beyond this modest door lies a marvelous world of dazzling complexity. 

Random tensors is a zero dimensional world, and, as such, it is background-independent; it makes no references whatsoever of any particular space-time. Moreover, based on the field theory of Feynman, it is  amenable  to renormalisation group techniques. Simple models even share with non-Abelian gauge theories the  property of asymptotic safety \cite{AF}.  
  
However until recently the lack of simple solvable examples of this correspondence prevented to extract easily the gravitational content. Gurau-Witten and Klebanov-Tarnoplosky built a first bridge \cite{GW} between tensors models and the gauge-gravity correspondance \cite{Mal} through conformal field theory, by making clever use of a modification of SYK models \cite{SYK}\footnote{We also point to the very recent review on the particular aspects of melonic CFTs that were discussed at the same conference \cite{Benedetti:2020seh}.}. But these models still are 
quantum mechanical and lost background-independence since they make use of a preferred time.

This paper is a very small step to restore
background-independence.
Zero dimensional tensor models create naturally trees or unicycles as Gromov-Hausdorff
limits. If we can aproximate the sub-dominant terms as matter fields living on
trees or unicycle (and it's a big ``if''), we shall get in this approximation an SYK-type model on a random tree or unicycle.
Thermal Euclidean, which plays such a natural role in SYK models, 
leads us to use unicycles rather than trees.
Models of this type can be studied
first by perturbative field theory techniques, hence the link with the paper \cite{DeRi}.

Then our main result is that, under reasonable assumptions, the SYK model for bosons averaged on (long, infrared) unicycles possesses a 
two-point function exhibiting  much the same behavior, but with a 
critical infrared exponent \emph{different} from the one of ordinary SYK, sensitive to the \emph{spectral dimension} of the underlying graph.

This can be seen as a simpler version of the well-known
2d CFT coupled to gravity (ie CFT on $\mathbb{R}^2$ but coupled to the Liouville field). The change in critical
exponent is a simpler analog of the Knizhnik-Polyakov-Zamolodchikov and David-Distler-Kawai relations, which tell how 
critical exponents change when coupled to 2d gravity.
The cycle in a unicycle can be identified to a (lattice regularized) flat $U(1)$ thermal circle, and the 
trees decorating the unicycles are then the unidimensional analog of the bidimensional Liouville field 
bumps which are at the source of the modification of critical exponents.
In this way, field theory  on random unicycles can be seen as "gravity in one dimension" or "gravitational time".

Relevance of the spectral dimension regarding renormalization properties of a quantum field theory on fractal sets was argued using a heat-kernel representation of the two-point function and resolved for hierarchical models, which allowed a rigorous Wilson block renormalization construction, by Eyink \cite{Eyink}.

This paper suggests research in many directions among which:
\begin{itemize}
\item understand the four point SYK function on unicycle. Does it still saturates MSS bound?
\item determine the renormalization group flow, CFT spectrum etc. in the lines of \cite{Benedetti_2019};
\item investigate the potential bulk holographic dual. Is it a kind of "random AdS$_2$"?
\item is there an analog of OS positivity for random spaces?
\end{itemize}

The text is organised as follows. In Section \ref{PRRT}, we review the material in \cite{DeRi}: the measure over the ensemble of trees, the field theory on the trees and relying on estimates over heat-kernels, perturbative bounds over amplitudes. In Section \ref{unicycles}, we introduce the unicycles as a compactified version of the random trees considered previously. We then recall very briefly the SYK model and its associated two-point function solution of a melonic Schwinger-Dyson equation. The next Section \ref{fermions} defines the analog of the SYK model on the unicycles. Finally, Section \ref{bosons} closes with the simpler extraction of the critical two-point function for a bosonic model on unicycles.

\section{Perturbative Renormalization on Random Trees}
\label{PRRT}

A graph $\Gamma$ is made of a set of vertices $V(\Gamma)$, a set of edges $E(\Gamma)$, 
and an incidence relation $R$ between them.
Fermions are preferably defined on oriented graphs (or digraphs), in which every edge has an arrow hence starts from a vertex $u$
and arrives at a vertex $v$. This will be noted as $e: u \to v$. Tadpoles $e: u \to u$ are a priori allowed although they will not be important in this paper.

\subsection{Critical Trees}

The Aldous continuous random tree \cite{aldous}
is the Gromov-Hausdorff limit of a critical Galton-Watson tree process with fixed branching rate conditioned on infinite survival along a single \emph{infinite spine} \cite{harris}.
One can also consider a discretized version of this continuous tree. It is a set of infinite discrete trees which have in common a unique infinite spine decorated with a product of independent measures for the finite critical Galton-Watson trees branching along the spine \cite{DJW}. We briefly recall the definition of the corresponding probability measure following closely \cite{DJW}.

The order $\vert T \vert$ of a rooted tree is defined as its number of edges, or its number of vertices different from the root $r$ (assumed, unless otherwise stated, to be of degree 1).
To a set of non-negative {\it branching weights} $w_i,\,i\in\mathbb
N^\star$ is associated the weights generating function $g(z):= \sum_{i \ge 1}  w_i z^{i-1} $ and the
{\it finite volume partition function} 
$Z_n$ on the set $\cT_n$ of all rooted trees $T$ of order $\vert T \vert =n$
\beq
Z_n = \sum_{T \in \cT_n}\prod_{u\in T \setminus r}w_{d_u}\,,
\eeq
where  $d_u$ denotes the degree of the vertex $u$.  The generating function for all $Z_n$'s
is 
\beq Z(\zeta ) = \sum_{n=1}^\infty Z_n \zeta^n ,
\eeq and it satisfies the equation
\beq\label{fixZ}
Z(\zeta ) =\zeta g(Z (\zeta)).
\eeq
Assuming a finite radius of convergence $\zeta_0$ for $Z$ one defines 
\beq
Z_0 = \lim_{\zeta\uparrow \zeta_0}Z(\zeta ) .
\eeq
The critical Galton-Watson
probabilities $p_i := \zeta _0 w_{i+1}Z_0^{i-1}$ for $i\in\mathbb N$ are then normalized: $\sum_{i=0}^\infty p_i = 1$. 
Any infinite tree in the class we consider has a root $r$, an \emph{infinite} spine $s_0 = r , s_1, s_2, \cdots , s_k , \cdots $, $k\in \mathbb N$,
plus a collection of $d_k -2$ \emph{finite} branches $T^{(1)}_k, \cdots , T^{(d_k -2)}_k$, at each vertex $s_k$
of the spine (recall the degree of $k$ is indeed $d_k$). 
The set of such trees is called $\cT_\infty$. It is equipped with a probability measure $\nu$ that we now describe.
This measure is obtained as a limit of measures $\nu_n$ on finite trees of order $n$. These measures $\nu_n$
are defined by \emph{identically and independently} distributing branches around a spine with measures 
\beq  \mu ( T) = Z_0^{-1}\zeta_0^{|T|}\prod_{u \in T \setminus r}w_{d_u}  = \prod_{i\in T\setminus r}p_{d_u -1}  \,.
\eeq

\noindent{\bf Theorem } \cite{DJW}
Viewing $\nu_n(T) = Z_n^{-1}\prod_{u\in T \setminus r}w_{d_u}\,,\quad\tau\in\cT_n\,,$ as a probability measure on $\cT$ we have
\beq
\nu_n \to \nu\quad \text{as}\quad n\to\infty\,,
\eeq
where $\nu$ is the probability measure on $\cT$ concentrated on the
subset $\cT_\infty$. 
\medskip

On random or fractal spaces, different notions of dimensions are introduced, in particular the Hausdorff and spectral dimensions. The first gives a global picture, close to the topological dimension. Formally, given a set $A$, its Hausdorff dimension is: \be d_H(A) = \inf\left\{d\geq 0: \lim_{r\to 0}\inf \left(\sum_i r_i^d\right)=0, \text{such that balls $S$ of radii $r_i\in (0,r)$ cover $A$}\right\}.\ee 
To the contrary, the spectral dimension provides a more local picture of the landscape. Given a set $A$ and a random walk $\{X_t\}_{t\geq 0}$ starting at $x\in A$, noting the probability that the random walker is at $y\in A$ at time $t$ by $p_t(y,x)$, the spectral dimension of $A$ is \be d_s(A) = -\f{1}{2}\lim_{t\to \infty}\f{\log p_t(x,x)}{\log t}. \ee 

\noindent{\bf Theorem }For generic infinite tree ensembles, the Hausdorff and spectral dimensions are respectively
$d_H=2$ and $d_{s} = 4/3\,$.

\medskip

For simplicity and in order not distract the reader's attention into unessential details we shall as in \cite{DeRi}
restrict ourselves from now on to the case of critical \emph{binary} Galton-Watson trees. 
It correspond to weights $w_1 = w_3 = 1$, and $w_i = 0$ for all other values of $i$. 
In this case the above formulas simplify. 
The critical Galton-Watson process corresponds to offspring probabilities
$p_0 =  p_2 = \frac{1}{2} $,  $p_i = 0$ for $i \ne 0, 2$.
The generating function for the 
branching weights is simply $g(z) = 1 + z^2$ and the generating function for the
finite volume trees $Z(\zeta ) = \sum_{n=1}^\infty Z_n \zeta^n  $
obeys the simple equation $
Z(\zeta ) =\zeta (1 + Z^2 (\zeta))$,
which solves to the Catalan function $Z= \frac{1 - \sqrt{1 - 4 \zeta^2}}{2\zeta}$. In the above notations 
the radius of convergence of this function is $\zeta_0 = \frac{1}{2}$. Moreover $Z_0 = \lim_{\zeta\uparrow \zeta_0}Z(\zeta ) =1$ and the measure on each branch of the Aldous tree is simply
\beq
\mu (T) = 2^{- \vert T \vert }\,.
\eeq

\subsection{Regularized Laplacians}

On any graph $\Gamma$, there is a natural notion of the Laplace operator $L_\Gamma$.
For a directed graph  $\Gamma$
the \emph{incidence matrix}  is the rectangular $V$ by $E$ matrix 
with indices running over vertices and
edges respectively, such that 
\begin{itemize}
\item
${\epsilon_\Gamma(v,e)}$ is +1 
if $e$  ends at $v$, 
\item
${\epsilon_\Gamma(v,e)}$ is -1 if $e$ starts at $v$,
\item
${\epsilon_\Gamma(v,e)}$ is  0 otherwise.
\end{itemize}

The $V $ by $V$ square matrix 
with entries  $d(v)$ on the diagonal is called the {\emph{degree matrix}}
$D_\Gamma$. 
The {\emph{adjacency matrix}} is the symmetric $V \times V$ matrix $A_\Gamma$ made of zeroes on the diagonal: $A_\Gamma(v,v) = 0 \;\; \forall v\in V$, and such that 
if $v \ne w$ then $A_\Gamma(v,w)$  is the number of edges of $G$ which have vertices $v$ and $w$ as their ends. 
Finally the {\emph{Laplacian matrix}} of $\Gamma$ is defined to be $\Delta_\Gamma = D_\Gamma - A_\Gamma$.
Its positivity properties stem from the important fact that it is a kind of square of the incidence matrix,
namely  
\bee
L_\Gamma = \epsilon_\Gamma \cdot \epsilon_\Gamma^\star.\ee
Remark that in graph theory the Laplacian is a positive rather than a negative operator, with a kernel dimension equal
to the number of connected components of $G$. So the sign convention for the graph Laplacian is the opposite of
the one of differential geometry.

The inverse $G_\Gamma$ of this operator  is formally given by the sum over random paths $\omega$ 
\beq  L_\Gamma^{-1} = G_\Gamma (x, y) = \sum_{\omega: x \to y} \; \prod_{v\in \Gamma} \;\biggl[\frac{1}{d_v}\biggr]^{n_v (\omega)}
\label{pathrep}
\eeq
where $d_v$ is the coordination at $v$ and $n_v (\omega)$ is the number of visits of $\omega$ at $v$.

As we know this series is not convergent without an infrared regulator, since a Laplacian always has
a constant zero mode. We can take out the zero mode by fixing a root vertex in the graph and deleting 
the corresponding line and column in $L_\Gamma$.
But to keep contact with the path representation \eqref{pathrep}, which has nice positivity properties, 
we shall use particular infrared regularizations compatible with this representation. 
There are two such natural regularizations.
\begin{itemize}
\item The mass regularization adds $m^2 \bbbone$ to the Laplacian, and leads to the path representation
\bee  G_\Gamma^m (x, y) = \sum_{\omega: x \to y} \;  \prod_{v\in \Gamma} \;\biggl[\frac{1}{d_v + m^2}\biggr]^{n_v (\omega)}
\label{masslap}
\ee
whose infrared limit corresponds to $m \to 0$.
\item The path length regularization (used in \cite{DJW}) corresponds to the path representation
\beq  G_\Gamma^\xi (x, y) = \sum_{\omega: x \to y} (1- \xi)^{\vert \omega\vert}  \; \prod_{v\in \Gamma} \;\biggl[\frac{1}{d_v}\biggr]^{n_v (\omega)}
\eeq
convergent for $0<\xi <1$; the infrared limit corresponds to $\xi \to 0$.
\end{itemize}

These two representations are usually equivalent for most infrared problems. For instance in our binary case
the path length representation  with constant $\xi$ corresponds to a mass regularization with just two different masses
$m^2_1 =  \frac{\xi}{1- \xi }$ on vertices of degree 1 and $m^2_3 =   \frac{3\xi}{1- \xi }= 3m^2_1$
on vertices of degree 3. The case of a regular lattice such as ${\mathbb Z}^d$ is even simpler; 
since its vertices have all constant degree $2d$,
the relationship is simply $m^2_d =  \frac{2d\xi}{1- \xi }$.

We want also to mimic fractional powers of the inverse Laplacian. This allows to probe for 
theories with scaling dimension  different 
from the one of the ordinary Laplacian. This is most conveniently done (again to respect the positivity
properties of the path representation) by superposing regularized Laplacians with a 
weight which diverges near $m^2 =0$ or $\xi =1$. For instance
in the case of the mass regularization we could define, for $0 < \alpha <1$, a kind of  K\"allen-Lehmann superposition
\beq  G_\Gamma^{\alpha} (x, y) = 2\int_0^\infty  m^{-2\alpha + 1}  G_\Gamma^m (x, y) dm .
\eeq

In the usual case on flat space it is easy to see why this corresponds to a fractional power
of the Laplacian. We have for example in the mass regularized case
the parametric representation 
\bee G^m (p) =(p^2 + m^2 )^{-1} = \int_0^\infty e^{-t (p^2 + m^2)} dt
\ee 
of the Laplacian in Fourier space. The corresponding superposed integrals are then
\bea
G^{\alpha} = (p^2 + m^2 )_\alpha^{-1} &=& \int_0^\infty m^{-2\alpha } 2mdm \int_0^\infty e^{-t (p^2 + m^2)}  dt \\
&=& \int_0^\infty e^{-u}  u^{-\alpha} du \int_0^\infty e^{-t p^2}  t^{-1 + \alpha} dt \\
&=& \Gamma (1- \alpha) p^{-2 \alpha} \int_0^\infty v^\alpha e^{-v} \frac{dv}{v}\\
&=& \Gamma (1- \alpha) \Gamma ( \alpha ) p^{-2 \alpha}  
\eea
where in the second line we put $u = tm^2$ and need $\alpha < 1$ for convergence of the $u$ integral near $0$; in the third line we put $v = tp^2$ and need $\alpha > 0$ for convergence of the $v$ integral near $0$. This shows that $G^{\alpha} $
is a fractional inverse Laplacian (with better infrared properties for $\alpha<1$ than the regular inverse Laplacian $p^{-2}$).

\subsection{Probabilistic Estimates}

The difficulty is to go from bounds valid on a single graph (quenched), to those for which an average over the ensemble of random graphs is taken (annealed). We based our analysis on the probabilistic estimates on random trees from
\cite{BarlowKumagai}.
More precisely we use the definition of $\lambda$-good balls, typical regions on the statistical ensemble of spaces we consider. We write the ball of radius $r$ from center $x$ as $B(x,r) = \{v\in \Gamma: d(x,v)\leq r\}$ and its volume $V(x,r)$ will count the number of vertices in $B(x,r)$. \footnote{In the notations of \cite{BarlowKumagai} and the random graph community, $V(x,r)$ sums, for all vertices $v$ in the ball $B(x,r)$, the weights associated to all edges attached to $v$. Since here the weights are all set to one and the graph is a tree, both definitions are proportional.}
For $\lam\ge 64$,
the ball $B(x,r)$ is said {\sl $\lam$--good} (Definition 2.11 of \cite{BarlowKumagai}) if:
\bea
r^2\lam^{-2} &\le& V(x,r) \le r^2\lam  , \label{goodvol} \\
m(x,r) &\le& \frac{1}{64} \lam, \quad
V(x, r/\lam) \ge r^2 \lam^{-4}, \quad V(x, r/\lam^2) \ge r^2 \lam^{-6}.
\eea
Remark that if $B(x,r)$ is {\sl $\lam$--good} for some $\lambda$,
it is {\sl $\lam'$--good} for all $\lambda' > \lambda$.

Corollary 2.12 of \cite{BarlowKumagai} proves that there are two constants $c_1,c_2$ such that:
\bee\mP  (B(x,r) \hbox{ is not $\lam$--good}) \le c_1 e^{-c_2 \lam}. 
\label{smallproba1}
\ee

Assuming that a random walk evolves in some $\lam$-good ball $B(x,r)$, essential quenched estimates are upper and lower bounds on the off-diagonal heat-kernel $q_{2t}(x,y)$ for a proper time $t$ restricted to an interval scaling as $r^3$.
\begin{theorem}
\label{theoBK1}
Suppose that $B=B(x,r)$ is $\lam$--good for 
$\lambda\ge 64$, and let  $I(\lam,r)=[r^3 \lam^{-6}, r^3 \lam^{-5}]$. Then 

\begin{itemize}
\item 
for any $K \ge 0$ and any $y\in T$ with $d(x,y) \le  Kt^{1/3}$
\bee  q_{2t}(x,y) \le  c\left(1+{\sqrt K}\right) t^{-2/3} \lam^{3}
\quad \hbox{\rm for } t\in I(\lambda, r) \;,  \label{gooddecay1}
\ee

\item
for any $y\in T$ with $d(x,y) \leq c_2 r \lambda^{-19}$
\bee  q_{2t}(x,y) \ge  c t^{-2/3} \lam^{-17}
\quad \hbox{\rm for } t\in I(\lambda, r).  \label{gooddecay2}
\ee

\end{itemize}

\end{theorem}

The theorem can be very useful.
In the multiscale decomposition of Feynman amplitudes, we introduce
$I_j = [ M^{2(j-1)}, M^{2j}] $ where $M$ has the dimension of mass, and we have the infrared equivalent continuous time representation
\bee C^{\alpha,j}_T (x,y) = \int_0^\infty u^{-\alpha } du \int_{I_j} q_t (x,y) e^{-ut } dt = \Gamma(1- \alpha)\int_{I_j} q_t (x,y)t^{\alpha -1} dt .
\label{trans}
\ee
Regarding the $\phi^4$ theory, for which $\alpha=\frac{1}{3} $ is the critical dimension such that the interaction is marginal or the theory just renormalizable, we used the above estimates to show upper and lower bounds on fractional annealed propagators (Lemma 3.3 in \cite{DeRi}).
\begin{lemma} 
\label{oneline}
There exist constants $c$, $c'$
such that the averaged propagator summed over one end-point obeys the bounds:
\bee   c M^{2j/3} \le \mE \left[ \sum_y C_T^{\alpha=\frac{1}{3} , j} (x,y) \right] \le c' M^{2j/3} .
\label{eq:lemma-singleline}
\ee
Also, there exists a third constant $\tilde{c}$ such that an annealed tadpole is bounded by:
\bee \mE \left[ C_T^j (x,x) \right] \le \tilde{c} M^{-2j/3} .
\label{cortadpole}
\ee
\end{lemma}
\proof The proof goes as follows. Given the scale $j$ and the point $x\in T$, we decompose the tree $T$ into annuli centered at $x$: 
\bee T = \cup_{k \in \N} \; A^T_{j,k}, \quad A^T_{j,k} := \left\{y : d(x,y) \in  [r_{j,k},r_{j,k+1}[\right\}, \label{sumT}
\ee
with the radii $r_{j,k} := M^{2j/3} k^{5/3}$. 
Taking $\lambda_{k,l}:= k+ l $ and $r_{j,k,l} := M^{2j/3}\lambda_{k,l}^{5/3}$, we are ensured that
\bee I_j = [M^{2j-2},M^{2j}] \subset I(\lambda_{k,l},r_{j,k,l})=
[r_{j,k,l}^3\lambda_{k,l}^{-6}, r_{j,k,l}^3\lambda_{k,l}^{-5}],
\ee
making Thm. \ref{theoBK1} applicable. In order to be able to sum over all points $y\in T$, we introduce the set of constants
$K_k := M^{2/3} (k+1)$ such that
\bee d(x,y ) \le K_k t^{1/3},\quad \forall t \in I_j,\;\forall y \in A^T_{j,k}. \label{distcond}
\ee
Finally, we also need to control the volume of balls $B^T_{j,k}$ of center $x$ and radius $r_{j,k}$ (or $B^T_{j,k,l}$ of radius $r_{j,k,l}$). In that purpose, we introduce the random variable $L=\min \{l \ge l_0 : B^T_{j,k,l} \text{ is } \lam_{k,l}\text{-good}\}$ allowing to partition the average over trees according to the volume of the ball $B_{j,k}$ containing $y$
\bee \mE \left[\sum_y C_T^j (x,y) \right] = \sum_{k=0}^\infty \sum_{l=l_0}^\infty \mP[L=l] \mE\vert_{L=l} \Bigl[\sum_{y \in A^T_{jk}} \int_{I_j} dt t^{\alpha-1} q_t (x,y)
\Bigr],
\ee
where $\mE\vert_{A}$ means conditional expectation with respect to the event $A$. Using the bounds \eqref{gooddecay1} and the fact that the volume of the annulus $A_{j,k}$ is bounded by that of the ball $B_{j,k,l}$, assumed $\lambda_{k,l}$-good but not $\lambda_{k,l+1}$-good, thus inequality \eqref{smallproba1}, we obtain the upper bound \eqref{eq:lemma-singleline}. The lower bound is easier since it is enough to restrict to the balls that are $\lambda_{k,l}$-good in the average over trees. Finally for the tadpole, the reasoning is straightforward since we are exempted of the sum over points in the balls.\qed

\noindent These estimates will enter the multiscale decomposition of Feynman amplitudes, through Kruskal greedy algorithm \cite{kruskal}.
It results in two theorems that establish the main {\it perturbative} bounds for 2PI graph:

\begin{theorem}\label{theoconv} For a completely superficially convergent graph $G$ (i.e. with no two- or four-point subgraphs), 
with vertex set $V(G)$ of order $n$, the limit as 
$\lim_{\rho \to \infty} \mE ( A_G)$
of the averaged amplitude exists 
and obeys the uniform bound
\bee \mE ( A_G) \le K^n (n!)^\beta \label{uniconv}
\ee
where $\beta =\frac{52}{3}$. 
\end{theorem}
An essential step in the proof is to bound from above the amplitude of the graph $G$ with that of a graph that keeps propagators connecting the vertices $V(G)$ with a (non-unique) Kruskal tree rooted at a vertex $x_0$, while the other propagators, from the loops, become tadpoles: 
\bee \mE [A_{G,\mu }] \le c^n \mE \Big[ 
 \sum_{\{x_{v} \}} 
\prod_{v=1}^{n-1} C_{T}^{j_v} (x_v, x_{a(v)})
\prod_{f =1}^{2n+2 - N} [C_T^{j_f}(x_f,x_f)]^{1/2} \Big] \; . \label{agmu3}
\ee
We have denoted by $a(v)$ the ancestor of the vertex $v$ assuming that the Kruskal tree directs the propagators pointing away from the root $x_0$. The remaining edges that formed loops split here into half-edges where tadpoles attach.

\begin{theorem}\label{theoconv1} For a graph $G$ with external points $N (G) \ge 4$, no two-point subgraph and with vertex set $V(G)$ of order $n$, the averaged effective-renormalized amplitude 
$\mE[ A_G^{eff}] =\lim_{\rho \to \infty} 
\mE[ A_{G,\rho}^{eff}]$ is convergent 
as $\rho \to \infty$ and obeys the same uniform bound as in the completely convergent case, namely
\bee \mE ( A_G^{eff}) \le K^n (n!)^\beta .
\ee
\end{theorem}
We refer to \cite{DeRi} for details of the proof. The value of $\beta$ has not being optimized.

Regarding now inclusion of two-point subgraphs, there is evidence from the long history of the critical behaviour of long-range models\footnote{See \cite{Behan:2017emf} for a comprehensive account.} that since divergences and the required counterterms are analytic functions of the external momenta, the fractional propagator is not renormalized  \cite{Honkonen1988fq, Slade_2017}.

\section{Unicycles}
\label{unicycles}

The cycle $\cC_\ell$ of length $\ell$ is the connected graph with $\ell$ vertices and $\ell$ edges forming a single circuit.
Unicyclic graphs $\Gamma$ are very mild modifications of trees. Instead of having \emph{no} cycle  they have a \emph{single cycle} $\cC(\Gamma) $. 
They can therefore be embedded on the sphere as planar graphs with \emph{two faces} 
(recall that trees have a \emph{single} ``external" face). 
The order $n= \vert \Gamma \vert$ of a unicycle $\Gamma$ 
is still defined as its total number of edges 
which is also its total number of vertices. 
Another important integer for a unicycle $\Gamma$ is its length $\ell  \ge 1$ which is defined as the
length of  $\cC(\Gamma) $. Hence $\ell \le n$.

\begin{figure}
\centerline{\includegraphics[width=10cm]{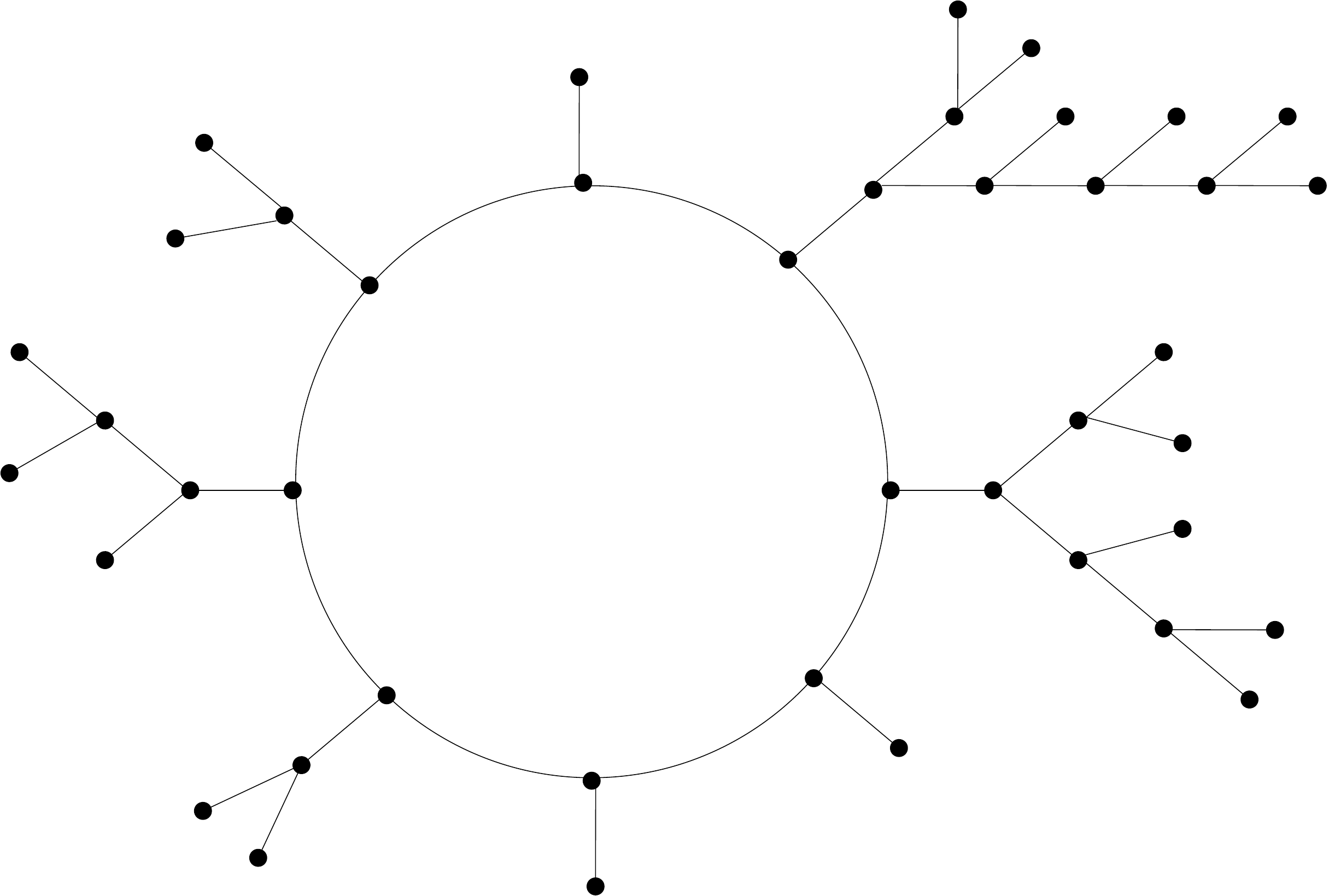}}
\label{unicyc}
\caption{This unicycle of length $\ell=8$ and order $n=42$ is binary: every vertex has degree either 3 or 1.}
\end{figure}
 
For simplicity we choose an orientation of the cycle and we orient every decorating tree from leaves to root, hence we 
can consider $\Gamma$ as an oriented graph or digraph.
Again for simplicity we shall restrict to the \emph{binary case}. It means 
that we shall consider unicycles whose vertices have degree either 1 or 3. 
All vertices of a cycle have degree $\ge 2$, hence in the binary case they must have degree three.
A binary unicycle of length $\ell$ is therefore 
characterized by the set of $\ell$ cyclically ordered rooted trees $T_0, \cdots , T_{\ell -1}$
attached to it.  It really means that the set $\cU_\ell$ of
unicyclic graphs of length $\ell$ can be identified to the set $[ \prod_{k=0}^{\ell -1} \cT_k]/{\rm Cyc}(\ell)$ of 
$\ell$ identical copies $\cT_k$ of the set of rooted binary trees $\cT$,
quotiented by the group ${\rm Cyc}(\ell)$ of cyclic permutations of $\{0, \cdots , \ell-1\}$.

Calling $C= \{t_0, \cdots t_{\ell-1} \}$ the set of vertices of the cycle,
since each $t_k$ is of degree 3, there is a set $\{t'_0, \cdots t'_{\ell-1}\}$ 
of tree vertices, each $t'_k$ being joined to $t_k$ by a single
edge not belonging to the cycle; therefore the order of 
a binary unicycle of length $\ell$ is at least $2\ell$ (see Figure \ref{unicyc}).

This characterization of binary unicycles suggests to define a probability measure for finite unicycles whose 
infinite order limit is closely related to the previous class $\cT_\infty$ of infinite binary random trees\footnote{Of course we expect that this limit is universal i.e. would be the same for $p$-ary random unicycles, but our emphasis here is not on this point.}.
The  cycle of the unicycle should be thought of as a finite analog of the \emph{spine} of an infinite tree in $\cT_\infty$.
Each $s_k$ is indeed the root of a binary Galton Watson tree
$T_k$, and we can equip these trees with independent probability measures $2^{- \vert T_k \vert}$.

We can then in the same vein as in \cite{DJW} define a measure $d\nu_\ell$ over the set $\cU_\ell$ of
unicyclic graphs of length $\ell$. It is just the product of independent 
Galton-Watson critical measures over the  attached trees $T_k$:
\beq  d \nu_\ell = \prod_{k=0}^{\ell -1}  \mu (T_k ).
\eeq
We shall not consider directly the limit $ \lim_{\ell \to \infty} d \nu_\ell$ since it is delicate to define
an analog of the \emph{infinite} spine \emph{with a periodic boundary condition}. Instead we can work 
with asymptotics of expectation values for $ d \nu_\ell$ as $\ell \to \infty$, just like
a thermodynamic or infrared limit can be defined as the large size limit of \emph{finite size} partition functions 
and correlation functions.

In the next section we shall therefore define the SYK model on unicycles in $\cU_\ell$ with finite $\ell$, 
define their correlation functions averaged over $d \nu_\ell$
and we shall study the infrared limit of these correlations when $\ell \to \infty$.
The cycle in the unicycle is the analog of
the lattice-regularized Euclidean time at a certain non-zero temperature. The trees of the unicycle
introduce the new "random gravity" aspects of this Euclidean time.

\section{The SYK Model}
\label{sec : syk}

In 2015 Kitaev proposed a very simple model \cite{SYK} which saturates the MSS bound \cite{MS2016}, indicating  the surprising presence of a gravitational dual in two dimensions. It is a disordered quantum mechanical model with action
\bee I = \int dt  \biggl( \frac{i}{2} \sum_{i=1}^N \psi_i (t) \frac{d}{dt}\psi_i (t)    
-i^{q/2} \sum_{1 \le i_1 <  \cdots < i_q \le N}	J_{i_1 \cdots  i_q}	\psi_{i_1}(t)  \cdots \psi_{i_q} (t)\biggr)
\ee
where $J_I$ is a quenched iid random tensor ($\expval{J_I J_{I'}} = \delta_{II'}J^2 (q-1)! N^{-(q-1)}$), and $\psi_i$ an $N$-vector Majorana Fermion.

This model now called the Sachdev-Ye-Kitaev model or \emph{SYK}
is solvable as $N \to \infty$, being approximately reparametrization invariant (i.e. conformal)
in the infrared limit. Moreover an attentive study of out-of-time ordered four-point functions reveals that the model
saturates the MSS bound \cite{MS2016}.
The corresponding so-called NAdS$_2$/NCFT$_1$ (where N stands for ``near") holography 
attracted considerable interest and is currently the subject of active investigation (see for example \cite{trunin2020} for a pedagogical exposition).

The model came somewhat as a surprise since solvability and chaotic behavior were previously somehow considered as incompatible. Since any near-extremal black hole should have a two dimensional "throat" 
in which the radial distance to the horizon and the time should be the effective interesting dimensions, the SYK model could 
shed light on the quantum aspects of black holes and in particular on hot issues such as the information paradox.

As mentioned above, the reason the theory can be \emph{solved}
in the limit $N \to \infty$ is because the leading Feynman graphs of the 
SYK model are the \emph{melonic} graphs \cite{Bon}
which dominate the $1/N$ tensor expansion \cite{1/N}. We give now 
a brief review of the corresponding computation of the two-point function in the infrared limit.

To start, we need to obtain the Euclidean two-point function 
\bee
G(\tau) = \frac{1}{N}\sum_i\expval{\psi_i(\tau)\psi_i(0)}
\ee
or (by abuse of notation) its Fourier transform
\bee
G(\omega) = \int_{-\infty}^\infty \dd{\tau} e^{i\omega \tau}G(\tau).
\ee
At finite temperature, the (Matsubara) frequencies are quantized: $\omega_n= \frac{2\pi }{\beta} (n + 1/2)$ and the Euclidean time is bounded $0\leq \tau\leq \beta$. It is convenient to define the quantity $\Delta = 1/q$.

\begin{figure}
\centerline{\includegraphics[height=3.5cm]{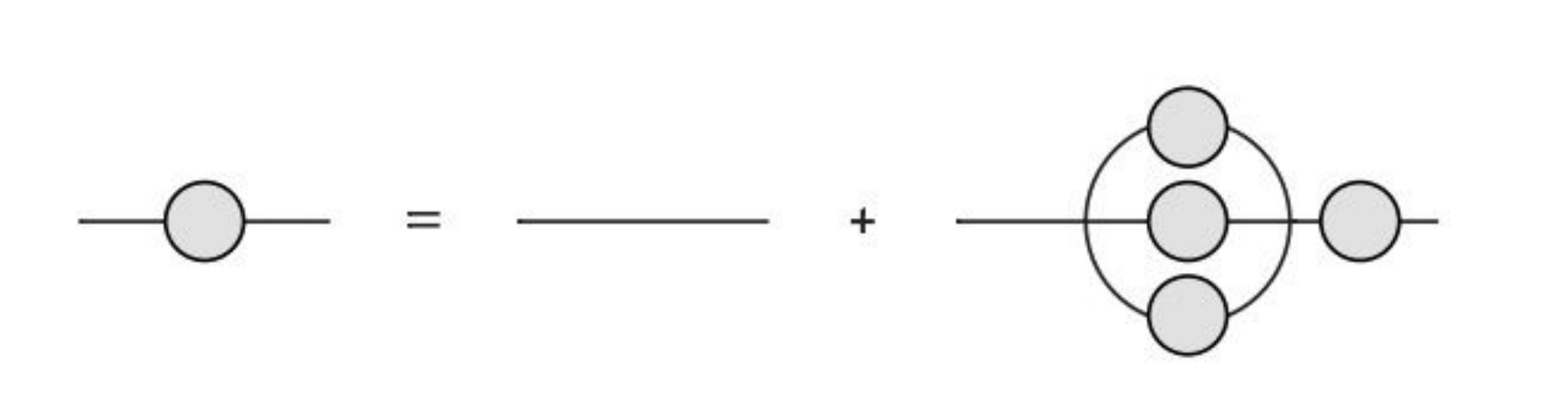}}
\caption{Melonic  Two-Point Function}
\label{fig : 2 point melonic}
\end{figure}

At leading order in $1/N$, after performing a quenched average
over $J$ and taking the limit $N \to \infty$, the Schwinger-Dyson equations for the free propagator $G_0^{-1} (\omega) = i \omega$ and the self-energy $\Sigma(\omega)$ simplify as 
\bee G^{-1} (\omega) = G_0^{-1} (\omega) - \Sigma (\omega),\quad  \Sigma (\tau) = J^2 G(\tau )^{q-1} .
\ee
This is pictured in Fig. \ref{fig : 2 point melonic}, where the ``blob" indicates a full propagator. 

\begin{figure}
\centerline{\includegraphics[height=4cm]{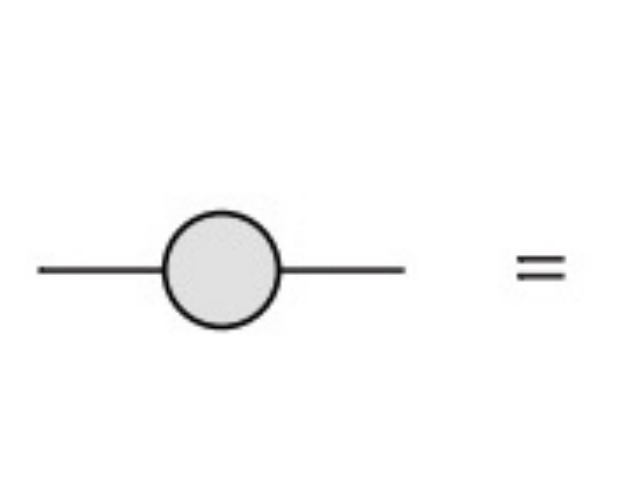}\includegraphics[height=4cm]{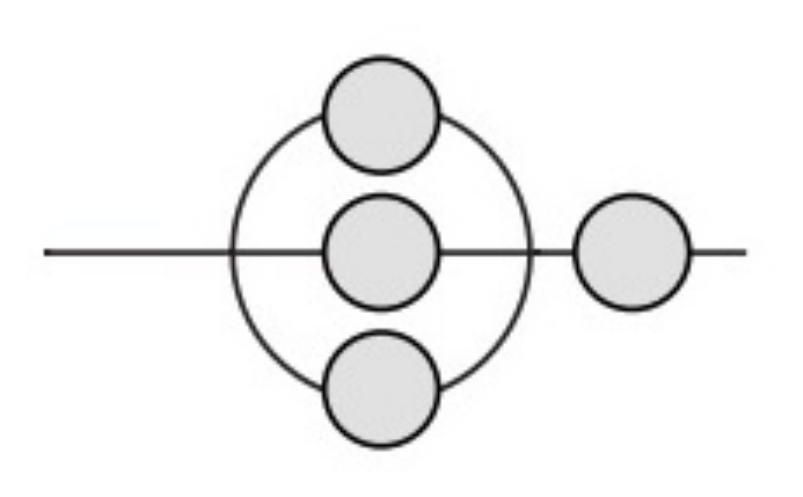}    }
\caption{Infrared Limit of Melonic Two-Point Function}
\label{fig : 2 point melonicIR}
\end{figure}

The first equation is the usual one linking the complete two-point function to the self-energy, the second is the melonic 
approximation which leads the $1/N$ tensor expansion.
Taking advantage of the form of the free propagator, in the IR limit the above equation simplifies to
the equation depicted in Fig. \ref{fig : 2 point melonicIR}, namely
\bee 
\label{eq:convolution}
\int \dd \tau^\prime J^2 G(\tau-\tau^\prime) G(\tau^\prime - \tau^{\prime \prime} )^{q-1} = -\delta (\tau - \tau^{\prime\prime}).
\ee

Reparametrization invariance of  \eqref{eq:convolution} under any differentiable function $f$:
\bea G (\tau, \tau') &\rightarrow& [ f'(\tau ) f'(\tau ') ]^\Delta  G(f(\tau), f(\tau ')), \\  
\Sigma (\tau, \tau') &\rightarrow& [ f'(\tau ) f'(\tau ') ]^{\Delta (q-1)}  \Sigma(f(\tau), f(\tau ')),
\eea
suggests to search for a particular solution of the type
\bee \label{2point}
G_c (\tau)  = b \vert \tau \vert^{-2 \Delta}  {\rm sign} \ \tau, \quad
J^2 b^q \pi = \left(\frac{1}{2} - \Delta\right) \tan (\pi \Delta ) .
\ee 
 The equation for $b$ comes from the formula
\bee \int_{- \infty}^{+ \infty} d \tau e^{i \omega \tau}  {\rm sign}  \tau \vert \tau \vert^{-2 \Delta} = 2^{1 - 2 \Delta} i \sqrt{\pi} \frac{\Gamma (1 - \Delta )}{  (\frac{1}{2} + \Delta)} \vert \omega \vert^{2 \Delta -1}  {\rm sign} \ \omega .
\ee
Applying reparametrization $f_\beta ( \tau) = \tan (\pi\tau/\beta )$ leads to 
the finite temperature two-point function
\bee G_c ( \tau) = \left[\frac{\pi }{\beta \sin (\pi \tau / \beta)} \right]^{2 \Delta} b \ {\rm sign } \;\tau.
\label{2pointfinitetemp}
\ee
Recalling that $\Delta = 1/q$, the anomalous dimension for $ G_c (\tau)  \propto  \tau^{-2/q}$ at large $\tau$
corresponds to the theory being just renormalizable in the  infrared (in the tensor field theory sense \cite{tensortrack}).

\section{Lattice-regularized SYK}
\label{fermions}

Consider a $\Gamma_0$ of length $\ell$ and order $n=l$, that is without any decoration by trees.
We want to define the {\it fermionic} SYK model on $\Gamma_0$. 
For the moment consider a one component Majorana fermion $\psi$, the generalization to $N$
components being straightforward.

The $U(1)$ Euclidean circle of length $\beta$
is replaced by an oriented finite cycle $C_\ell = \{t_0 , \cdots t_{\ell -1} \}$ with  $t_k = ak$
and $\ell = \beta/a$. The ultraviolet limit $a \to 0$
and infrared limit $\beta \to \infty$ then both imply $\ell \to \infty$, by keeping constant the perimeter of the circle.

We have first to implement the antiperiodic boundary conditions. Antiperiodicity on the lattice means that 
$\psi$ is in fact periodic but with period $2 \beta$ instead of $\beta$ hence should be analyzed in terms of 
$2 \ell$ frequencies $ \frac{\pi q}{\beta}  =  \frac{\pi q }{a \ell}  $ with $q = 1 , \cdots 2 \ell$, but that 
only odd Matsubara frequencies $\omega_p =(2p+1) \frac{\pi }{\beta} $ for $p = 0, \cdots \ell -1$ contribute.
The discrete Fourier transforms are defined as
\bee \hat \psi (\omega_p) = 
\frac{1}{\sqrt \ell} \sum_{k=0}^{\ell -1} e^{- i \omega_p \cdot t_k} \psi (t_k) = \frac{1}{\sqrt \ell} \sum_{k=0}^{\ell -1}
e^{- \frac{(2p+1)i \pi  k}{\ell} }  \psi (ak) .
\ee
Remark that 
\bee \hat \psi (\omega_{\ell -p -1}) = \overline{\hat \psi (\omega_p)} \label{compconj}
\ee
and that the inverse Fourier law gives antiperiodic fields of antiperiod $\beta$
\bee
\psi (t_k) =  \frac{1}{\sqrt \ell} \sum_{p=0}^{\ell -1}  
e^{i\omega_p \cdot t_k} \hat \psi ( \omega_p) = 
\frac{1}{\sqrt \ell} \sum_{p=0}^{\ell -1} e^{\frac{(2p+1) i \pi  k}{\ell} } \hat \psi \left( \frac{(2p+1)\pi }{\beta}  \right) = - \psi (t_k + \beta) .
\ee

The action should be discretized in the usual way, that is turning each derivative at $t_k$ into
a lattice difference 
\bee \frac{1}{a} [ \psi (ka +a) -\psi (ka) ] =  \frac{1}{a}  [ \psi (t_{k+1}) -\psi (t_k) ] 
\ee and integrals such as $\int dt f(t)$
into Riemann lattice sums $ a \sum_{k=0}^{\ell -1} f(t_k)$.
To discretize the quadratic part $ \frac{i}{2}   \int dt \sum_{i=1}^N \psi_i (t) \frac{d}{dt}\psi_i (t)   $ of the SYK action one
should take  into account the anticommutation of Fermions.
Factoring out $\frac{i}{2}$ leads to consider the quadratic form  
\bee Q_{lat} (\psi) = \,\left[ \,  
\sum_{k=0}^{\ell -2} \psi (t_k)  \psi (t_{k+1})   - \psi (t_{\ell-1}) \psi ( t_0) \right]  , \label{SYKlataction}
\ee 
where the last term is subtle: because of  antiperiodicity, $\psi(t_\ell)$ should be identified with $ - \psi (t_0)$.
The total interacting SYK lattice action is therefore
\bee I_{lat} =\sum_{i=1}^N  \frac{i}{2}  Q_{lat} (\psi_i) -a i^{q/2}  \sum_{k=0}^{\ell -1}
\sum_{1 \le i_1 <  \cdots < i_q \le N}	J_{i_1, \cdots , i_q}	\psi_{i_1}(t_k) \cdots \psi_{i_q} (t_k) .
\ee
Remark that in order to have a non zero normalization (for $q$ even) the total number $\ell$
should be even, a condition which we assume from now on.

We can rewrite the quadratic (free) action in Fourier space. Forgetting the trivial $i$ index, it means
\bea Q_{lat} (\psi) &=& \Bigl[ \,  
\sum_{k=0}^{\ell -2} \psi_i (t_k)  \psi (t_{k+1})   - \psi (t_{\ell-1}) \psi( t_0) \Bigr]  \\
&=& \frac{1}{\ell} \sum_{p=0}^{\ell -1}  \sum_{q=0}^{\ell -1} 
\biggl[\sum_{k=0}^{\ell -2}  e^{\frac{i \pi }{\ell} [(2p+1)k +(2q +1) (k+1)]} 
 -  e^{\frac{i \pi }{\ell} [(2p+1) (\ell -1) }  \Biggr]
 \hat \psi (\omega_p ) \hat \psi (\omega_q ) \nonumber \\
 &=& \frac{1}{\ell} \sum_{p=0}^{\ell -1}  \sum_{q=0}^{\ell -1} 
\biggl[ e^{\frac{i \pi (2q+1)}{\ell}}  \sum_{k=0}^{\ell -1}  e^{\frac{2i \pi }{\ell} (p+q +1) k} \Biggr]
 \hat \psi (\omega_p ) \hat \psi (\omega_q )\\
 &=&  \sum_{p=0}^{\ell -1} 
\biggl[ e^{- \frac{i \pi ( 2p +1)}{\ell}} -1   \Biggr]
 \hat \psi (\omega_p )  \overline{\hat \psi (\omega_p )}.
\eea 
where in the last line we took advantage of the fact that the sum over $k$ gives zero
unless $e^{\frac{2i \pi }{\ell} (p+q +1)} =1$ hence $p+q +1 = \ell$,
and we added the $-1$ because it is the Fourier transform of a mass term 
hence is zero, and we used \eqref{compconj} .

Remark  that the factor $e^{\frac{-i \pi ( 2p +1)}{\ell}}  - 1$ is never zero and behaves as $-i \pi (2p+1)$ for small $p$. Hence
the  free lattice propagator is invertible and approximates at small $p$ 
the inverse of the continuous free propagator, hence the 
inverse of the Matsubara frequency.

Consider now any unicycle $\Gamma$ with decorating trees
oriented from leaves to the cycle. We want to define the fermionic SYK model on $\Gamma$. 

We impose two conditions. First we want the free action of the $N$-component Majorana field $\psi$ to be a quadratic form 
$\frac{i}{2}  Q_\Gamma (\psi)$
with a good non-zero normalization. Second we want to impose
the anti-periodic boundary conditions along the spine to
coincide with the ones of the ordinary free SYK model on the trivial unicycle $\Gamma_0$ (not decorating by the trees).

A naive quadratic form would couple a Fermion on each vertex to its nearest neighbours. However it does not work, since as soon 
as a single branch of $\Gamma$ is non trivial, the corresponding free theory has zero normalization. 
Indeed in this case the tree 
has a non trivial terminal branch with two leaves $s_1$ and $s_2$ related to a node $s_3$, and the Grassmann integral contains a term such as  $\int d\psi (s_1 ) d \psi(s_2) 
d\psi(s_3)   e^{\psi(s_1) \psi(s_3)  + \psi(s_2) \psi(s_3)}$, which is zero.

In fact our two conditions lead to the same conclusion, namely that we need some kind of Fermion doubling. The SYK model on $\Gamma$ requires  a single $N$-component Fermion variable not only for the $n$ vertices of $\Gamma$
but also for the $n$ edges of $\Gamma$. 
With this convention we can define
\bee Q_{\Gamma} (\psi) = \,\Bigl[ \,  
\sum_{e,v} \epsilon_{ev} \psi (e)  \psi (v)  \Bigr]  .\label{SYKlataction}
\ee

To implement the anti-periodicity we fix a particular  root vertex $v_0$ on the spine  and
we reverse the last half-arrow $e_\ell $ into that vertex.

\begin{lemma}
The normalization  of the free theory at any unicycle $\Gamma$, $Z_{\Gamma}$, is $2^{N\ell}$, so that 
\bea
Z_{\Gamma}^{-1} \int e^{Q_{\Gamma} (\psi)} \prod d \psi = 1.
\eea
\end{lemma}
\proof First treat the spine, then by induction, adding a leaf. \qed

There are really $2\ell$ fermions on the spine, $\ell$ ones for the vertices and $\ell$ ones for the edges. 
This allows to automatically implement the evenness condition, hence we can use the normalized quadratic form $Q_C$
in the form
\bee Q_{lat} (\psi) = \,\Bigl[ \,  
\sum_{k=0}^{\ell -2} \psi (t_k)  \psi (t_{k+1})   - \psi (t_{\ell-1}) \psi ( t_0) \Bigr]  . \label{SYKlataction}
\ee 

Let us note that this is the most simple example of introducing fermions on a graph, the general case taking the Dirac operator as acting on the cliques of the graph \cite{Knill(2013)} \footnote{See also \cite{Gubser:2018cha} for a similar definition of fermions on a tree graph in the context of p-adic AdS/CFT.}.

\section{Bosons}
\label{bosons}

For the moment let us focus on the two-point function for the SYK in the simplest case of Bosons. The bare SYK model on 
$\Gamma$ can be defined by the discretized action
\bee I_\Gamma = \sum_{u \in V(\Gamma)} \biggl[ \frac{1}{2}\phi (u)  ( L_\Gamma + m^2 \bbbone) \phi (u) 
-\frac{i^{q/2}}{q!} \sum_{1 \le i_1 <  \cdots < i_q \le N}	J_{i_1 \cdots  i_q}	\phi_{i_1}(u) \cdots \phi_{i_q} (u) \biggr].
\ee
where $L_\Gamma$ is the lattice Laplacian on $\Gamma$ and the disordered coupling has second moment:
\bee
\expval{J_{i_1 \cdots  i_q}^2} = \frac{(q-1)!J^2}{N^{q-1}}. 
\ee 
In the $N \to \infty$ limit we get a self-consistent melonic equation for the two-point function $G^{mel} (x,y)$
\bee 
[G^{mel} (x,y)]^{-1}  = - [G_0^{mel} (x,y) ]^{-1} - \int \dd \tau^\prime J^2 [G^{mel}(x,y)]^{q-1} .
\label{eq:convolution1}
\ee
which simplifies in the infrared limit \cite{Gurau:2017qna}
into the convolution equation:
\bee 
\sum_{z \in \Gamma}  J^2 G_{ir}^{mel}(x,z)  [G_{ir}^{mel}(z,y)]^{q-1}= -\delta (x,y).
\label{eq:convolution2}
\ee
We shall now \emph{assume} that the effective infrared two point function $G^{mel}_{ir} (x,y)$  is 
asymptotic to  an $\alpha$-regularized propagator on $\Gamma$, namely
$ G_\Gamma^{\alpha} (x, y) = 2\int_0^\infty  m^{-2\alpha + 1}  G_\Gamma^m (x, y) dm $. We shall average
over unicycles $\Gamma$ of given length $\ell $ and search for the right value of $\alpha$
to fulfill \eqref{eq:convolution2}. Since we average over different $\Gamma$'s but which share all the same 
cycle of length $\ell$ it makes sense to consider $x$ and $y$ in that cycle, but the intermediate point $z$ 
can be anywhere on $\Gamma$, including in the decorating trees.

So we are searching for the value of $\alpha $ such that for $x, y \in \cC_\ell$ the equation
\bee 
\expval{\sum_{z \in \Gamma}  J^2 G_\Gamma^{\alpha} (x, z)  [G_\Gamma^{\alpha} (z, y) ]^{q-1}}_{\ell} = -\delta (x,y) .
\label{eq:convolution3}
\ee
holds. In this discrete setting, the right hand side is a Kronecker delta. Consequently we take $x \simeq y$ on the left and the Kruskal tree is made of a single propagator connecting $x$ to $z$.
The average $\expval{.}_\ell$ means  averaging with $d\nu_\ell$, hence over all unicycles of length $\ell$ with independent 
critical binary Galton-Watson 
measure on the trees decorating the cycle.

Then the multiscale analysis is especially interesting. 

According to the earlier Lemma~\ref{oneline}, it results in
\bee   \mE \left[ \sum_z C_T^{\alpha , j} (x,z) \right] \simeq M^{\frac{4j}{3}} M^{-\frac{4j}{3}+2\alpha j} ,
\label{singleline1}
\ee
and the tadpoles count each for 
\bee   \mE \left[  C_T^{\alpha , j} (x,x) \right] \simeq  M^{-\frac{4j}{3}+2\alpha j} .
\label{singleline2}
\ee
Remembering the proof of Theorem~\ref{theoconv} which can be framed as
\bee \mE  \left[ \left(\sum_z C_T^{\alpha , j} (x,z) \right) (  C_T^{\alpha , j} (x,z) )^{q-1} \right]  \simeq \mE \left[  C_T^{\alpha , j} (x,z) \right]\mE \left[ \sum_z C_T^{\alpha , j} (x,x) \right] ^{q-1} ,
\ee
and collecting all factors,
we have
\bee   \mE \left[ \left(\sum_z C_T^{\alpha , j} (x,z) \right) (  C_T^{\alpha , j} (x,x) )^{q-1} \right] \simeq  M^{-\frac{4(q-1)j}{3} +2q\alpha j} .
\label{singleline3}
\ee
Returning to eq.~\eqref{eq:convolution3}, we find
\bee   M^{-\frac{4(q-1)j}{3} +2q\alpha j} \simeq   M^{-\frac{4j}{3}}  \implies \alpha = \frac{2}{3}  - \frac{4}{3q}.
\label{singleline4}
\ee
Let us note that the scaling associated to the Kronecker delta that selects the critical value of $\alpha$ making the $q$-th order interaction marginal, must correspond to the spectral dimension (here $4/3$).

The effective infrared propagator, Fourier transforming on the spine, then behaves on the unicycle as
\bee
\expval{G^{\alpha}(p) }_\ell \simeq p^{-2 \alpha} = p^{-\frac{4}{3} +  \frac{8}{3q}} ,
\ee
hence very differently from the flat deterministic case $p^{-1+2/q}$. This is the main result of the paper and, as usual, it corresponds to
the just renormalizable case. Confirming this, for the moment crude, analysis of the melonic equation with finer heat-kernel estimates is an interesting problem we are working on.

\section*{Aknowledgments}
\noindent
Nicolas Delporte 
would like to thank D. Benedetti and J. Ben Geloun for various helpful discussions. 
Vincent Rivasseau thanks George Zoupanos for making such a wonderful series of workshops.


\begin{thebibliography}{99}

   \bibitem{DeRi}
  N.~Delporte and V.~Rivasseau,
  ``Perturbative Quantum Field Theory on Random Trees,''
 to appear  in Commun. Math. Phys; arXiv:1905.12783.
  
   \bibitem{tensortrack}
  V.~Rivasseau,
 ``Quantum Gravity and Renormalization: The Tensor Track,''
  AIP Conf.\ Proc.\ 1444, 18 (2011);
 % [arXiv:1112.5104 [hep-th]].
  ``The Tensor Track: an Update,''
  arXiv:1209.5284 [hep-th];
  ``The Tensor Track, III,''
  Fortsch.\ Phys.\  {\bf 62}, 81 (2014);
  %[arXiv:1311.1461 [hep-th]].
  ``The Tensor Track, IV,''
  PoS CORFU 2015, 106 (2016);
  %[arXiv:1604.07860 [hep-th]].
  ``Random Tensors and Quantum Gravity,''
  SIGMA {\bf 12}, 069 (2016).
 % [arXiv:1603.07278 [math-ph]].

\bibitem{tensors}
R.~Gurau, ``Random Tensors", Oxford University Press (2016);
R.~Gurau,  ``Invitation to Random Tensors,''
%  SIGMA {\bf 12}, 094 (2016)
% % doi:10.3842/SIGMA.2016.094
% [arXiv:1609.06439 [hep-th]],
in SIGMA special issue "Tensor Models, Formalism and Applications", 2016;
 R.~Gurau and J.~P.~Ryan, ``Colored Tensor Models - a review,''
SIGMA {\bf 8}, 020 (2012),
%  doi:10.3842/SIGMA.2012.020
%[arXiv:1109.4812].



\bibitem{Bon} 
V.~Bonzom, R.~Gurau, A.~Riello and V.~Rivasseau,
``Critical behavior of colored tensor models in the large N limit,''
Nucl.\ Phys.\ B {\bf 853}, 174 (2011);
  V.~Bonzom, R.~Gurau and V.~Rivasseau,
 ``Random tensor models in the large N limit: Uncoloring the colored tensor models,''
  Phys.\ Rev.\ D {\bf 85}, 084037 (2012)
%[arXiv:1105.3122 [hep-th]].
  
  
 

\bibitem{AF} 
  J.~Ben Geloun,
  ``Two and four-loop $\beta$-functions of rank 4 renormalizable tensor field theories,''
  Class.\ Quant.\ Grav.\  {\bf 29}, 235011 (2012);
    J.~Ben Geloun and D.~O.~Samary,
  ``3D Tensor Field Theory: Renormalization and One-loop $\beta$-functions,''
  Annales Henri Poincare {\bf 14}, 1599 (2013);
 % [arXiv:1205.5513 [hep-th]]; 
%
 % S.~Carrozza and V.~Lahoche,
  %``Asymptotic safety in three-dimensional SU(2) Group Field Theory: evidence in the local potential approximation,''
 % Class.\ Quant.\ Grav.\  {\bf 34}, no. 11, 115004 (2017)
 % [arXiv:1612.02452 [hep-th]].
   %  \cite{Rivasseau:2015ova}
%\bibitem{Rivasseau:2015ova} 
  V.~Rivasseau,
  ``Why are tensor field theories asymptotically free?''
  Europhys.\ Lett.\  {\bf 111}, no. 6, 60011 (2015)
  %[arXiv:1507.04190 [hep-th]].
   

%\bibitem{Seiberg:2006wf}
%N.~Seiberg,
%``Emergent space-time,''
%arXiv:hep-th/0601234.
  


 %\cite{Witten:2016iux}
\bibitem{GW} 
  E. Witten,
 ``An SYK-Like Model Without Disorder,''
  [arXiv:1610.09758 [hep-th]];
  %%CITATION = ARXIV:1610.09758;%%
  %133 citations counted in INSPIRE as of 04 Apr 2018
  R. Gurau,
  ``The complete $1/N$ expansion of a SYK-like tensor model,''
  Nucl.\ Phys.\ B {\bf 916}, 386 (2017);
  %doi:10.1016/j.nuclphysb.2017.01.015
 % [arXiv:1611.04032 [hep-th]];
  %%CITATION = doi:10.1016/j.nuclphysb.2017.01.015;%%
  %56 citations counted in INSPIRE as of 21 Feb 2018
  S. Carrozza and A.Tanasa,
  ``$O(N)$ Random Tensor Models,''
  Lett.\ Math.\ Phys.\  {\bf 106}, no. 11, 1531 (2016);
 % [arXiv:1512.06718 [math-ph]].
  %%CITATION = doi:10.1007/s11005-016-0879-x;%%
  %41 citations counted in INSPIRE as of 08 Mar 2018  
%\bibitem{Klebanov:2016xxf}
I. Klebanov and G. Tarnopolsky, ``Uncolored Random Tensors, Melon
  Diagrams, and the SYK Models,"  Phys. Rev. {\bf D95} (2017) 046004
 % [arXiv:1611.08915 [hep-th]].
%%CITATION = ARXIV:1611.08915;%%

  \bibitem{Mal}
J. Maldacena, ``The Large N limit of superconformal field theories and supergravity", Advances in Theoretical and Mathematical Physics, vol. 2, 1998, p. 231-252;
O. Aharony, S. Gubser, J. Maldacena, H. Ooguri, Y. Oz, "Large N Field Theories, String Theory and Gravity". Phys. Rep. 323 (3-4): 183-386. 
%[arXiv:hep-th/9905111];
  N.~Beisert {\it et al.},
  ``Review of AdS/CFT Integrability: An Overview,''
  Lett.\ Math.\ Phys.\  {\bf 99}, 3 (2012) 
  %[arXiv:1012.3982 [hep-th]].
 %\cite{BenGeloun:2012pu}


%\cite{Maldacena:2015waa}
\bibitem{SYK} 
A.~Kitaev, ``A simple model of quantum holography.''
Talks at KITP, April 7, 2015 and May 27, 2015;
  J.~Polchinski and V.~Rosenhaus,
  ``The Spectrum in the Sachdev-Ye-Kitaev Model,''
  JHEP {\bf 1604}, 001 (2016);
  %doi:10.1007/JHEP04(2016)001
%  [arXiv:1601.06768 [hep-th]].  
%\cite{MS2016}
% \bibitem{MS2016} 
J.~Maldacena and D.~Stanford,
  ``Remarks on the Sachdev-Ye-Kitaev model,''
  Phys.\ Rev.\ D {\bf 94}, no. 10, 106002 (2016).
  %doi:10.1103/PhysRevD.94.106002
 % [arXiv:1604.07818 [hep-th]]

%\cite{Benedetti:2020seh}
\bibitem{Benedetti:2020seh}
D.~Benedetti,
``Melonic CFTs,''
[arXiv:2004.08616 [hep-th]].
%0 citations counted in INSPIRE as of 28 Apr 2020
 
 \bibitem{Benedetti_2019}
   Benedetti, Dario and Gurau, Razvan and Harribey, Sabine, ``Line of fixed points in a bosonic tensor model",
   JHEP {\bf 06} (2019) 053,
   [arXiv:1903.03578]
   %ISSN={1029-8479},
   %url={http://dx.doi.org/10.1007/JHEP06(2019)053},
   %DOI={10.1007/jhep06(2019)053},
   %number={6},
   
%\cite{Eyink:1989dv}
\bibitem{Eyink}
G.~Eyink,
``Quantum Field Theory Models on Fractal Space-time. 1: Introduction and Overview,''
Commun. Math. Phys. \textbf{125} (1989), 613-636
%doi:10.1007/BF01228344
 
 %\cite{Maldacena:2015waa}
\bibitem{MS2016}
  J.~Maldacena, S.~H.~Shenker and D.~Stanford,
  ``A bound on chaos,''
  JHEP {\bf 1608} (2016) 106
  %doi:10.1007/JHEP08(2016)106
  [arXiv:1503.01409 [hep-th]].
%   %%CITATION = doi:10.1007/JHEP08(2016)106;%%
%   %779 citations counted in INSPIRE as of 06 Apr 2020

\bibitem{trunin2020}
  D.~A.~Trunin,
  ``Pedagogical introduction to SYK model and 2D Dilaton Gravity,''
  [arXiv:2002.12187 [hep-th]].


\bibitem{aldous} D. Aldous, ``The Continuum Random Tree I, II and III", The Annals of Probability, 1991, Vol 19, 1-28; in 
Stochastic Analysis, London Math Society Lecture Notes, Cambridge University Press 1991n eds Barlow and Bingham;
The Annals of Probability, 1993, Vol 21, 248-289.

\bibitem{harris}T.~E.~Harris, {\it The theory of branching processes,}
Dover Publiations, Inc., New York (2002)


\bibitem{DJW} B. Durhuus, T. Jonsson and J. F. Wheater, {\it The Spectral Dimension of Generic Trees,}  Journal of Statistical Physics
{\bf 128}, Issue 5, 1237-1260 (2007)

\bibitem{BarlowKumagai}
M. T. Barlow and T. Kumagai, {\it Random Walk on the Incipient Infinite
cluster on trees}, Illinois Journal of Mathematics, Vol. 50, 33-65 (2006).

%\cite{Behan:2017emf}
\bibitem{Behan:2017emf}
C.~Behan, L.~Rastelli, S.~Rychkov and B.~Zan,
``A scaling theory for the long-range to short-range crossover and an infrared duality,''
J. Phys. A \textbf{50} (2017) no.35, 354002,
%doi:10.1088/1751-8121/aa8099
[arXiv:1703.05325 [hep-th]].
%42 citations counted in INSPIRE as of 28 Apr 2020

%\cite{Honkonen:1988fq}
\bibitem{Honkonen1988fq}
J.~Honkonen and M.~Nalimov,
``Crossover between field theories with short range and long range exchange or correlations,''
J. Phys. A \textbf{22} (1989), 751-763.
%doi:10.1088/0305-4470/22/6/024
%15 citations counted in INSPIRE as of 28 Apr 2020

\bibitem{Slade_2017} Lohmann, M., Slade, G., Wallace, B.~C.\ ``Critical Two-Point Function for Long-Range $O(n)$ Models Below the Upper Critical Dimension,"\ Journal of Statistical Physics 169, 1132,  [arXiv:1705.08540 [math-ph]]


\bibitem{kruskal}
J.~B.~Kruskal,
{\it On the Shortest Spanning Subtree of a Graph and the Traveling
  Salesman Problem},
Proceedings of the American Mathematical Society 7, 48 (1956).

\bibitem{1/N}
  R.~Gurau,
 ``The 1/N expansion of colored tensor models,''
  Annales Henri Poincar\'e {\bf 12}, 829 (2011);
  R.~Gurau and V.~Rivasseau, ``The 1/N expansion of colored tensor models in arbitrary dimension,''
  EPL {\bf 95}, no. 5, 50004 (2011);
  R.~Gurau,
  ``The complete 1/N expansion of colored tensor models in arbitrary dimension,''
  Annales Henri Poincar\'e {\bf 13}, 399 (2012);


  %%CITATION = doi:10.1007/s11005-011-0529-2;%%
  %764 citations counted in INSPIRE as of 04 Apr 2018
%\cite{Gurau:2017qna}
\bibitem{Gurau:2017qna} 
  R.~Gurau,
  ``The $\imath \epsilon$ prescription in the SYK model,'' J. Phys. Comm.2 (2018) 015003.
  %%CITATION = ARXIV:1705.08581;%%
  %19 citations counted in INSPIRE as of 11 Feb 2019
  
  
\bibitem{Knill(2013)} O. Knill,\  ``The Dirac operator of a graph",\ [arXiv:1306.2166]. 
 %\cite{Gubser:2018cha}
\bibitem{Gubser:2018cha}
S.~S.~Gubser, C.~Jepsen and B.~Trundy,
``Spin in $p$-adic AdS/CFT,''
J. Phys. A \textbf{52} (2019) no.14, 144004
%doi:10.1088/1751-8121/ab0757
[arXiv:1811.02538 [hep-th]].
%17 citations counted in INSPIRE as of 28 Apr 2020



\end{thebibliography}
\end{document}